\documentclass[11pt]{article}
\usepackage{graphics}
\usepackage{latexsym}
\usepackage{extra}

\begin{document}

\title{Greedy Facility Location Algorithms \\Analyzed using Dual Fitting
with Factor-Revealing LP }
\author{
{Kamal Jain}\thanks{Microsoft Research, One Microsoft Way,
Redmond, WA 98052. E-mail: {\sf kamalj@microsoft.com}}\hspace{8mm}
\and
{Mohammad Mahdian}\thanks{Department of Mathematics, MIT, Cambridge, MA 02139, USA. E-mail: {\sf mahdian@math.mit.edu}.}\hspace{8mm} \and 
\and {Evangelos Markakis}\footnotemark[3]\hspace{8mm}
\and {Amin Saberi}\footnotemark[3]\hspace{5mm}
\and {Vijay V. Vazirani}\thanks{College of Computing, 
Georgia Tech, Atlanta, GA 30332, USA. E-mails: {\sf
\{vangelis, saberi, vazirani\}@cc.gatech.edu}.} }
\date{}
\maketitle

\begin{abstract}
In this paper, we will formalize the method of dual fitting and the idea 
of factor-revealing LP. This combination is used to design and analyze
two greedy algorithms for the metric uncapacitated facility 
location problem. Their approximation factors are 1.861 and 1.61, with 
running times of $O(m\log m)$ and $O(n^3)$, respectively, where
$n$ is the total number of vertices and $m$ is the number of edges 
in the underlying
complete bipartite graph between cities and facilities.
The algorithms are used to improve recent results for several
variants of the problem.
\end{abstract}

\section{Introduction}

A large fraction of the theory of approximation algorithms, as we
know it today, is built around the theory of linear programming,
which offers the two fundamental algorithm design techniques of
rounding and the primal--dual schema (see~\cite{Va.book}). 
Interestingly enough, the LP-duality based analysis~\cite{lovasz,chvatal} 
for perhaps the most central problem of this theory, 
the set cover problem, did not use either of these techniques.
Moreover, the analysis used for set cover does not seem to have found use outside 
of this problem and its generalizations~\cite{multicover}, leading to
a somewhat unsatisfactory state of affairs.

In this paper\footnote{This paper is based on the preliminary versions
\cite{MMSV} and~\cite{JMS}.},
we formalize the technique used for analyzing set cover as the
{\em method of dual fitting}, and we also introduce the idea 
of using a {\em factor-revealing LP}. Using this combination we analyze
two greedy algorithms for the metric uncapacitated facility 
location problem. Their approximation factors are 1.861 and 1.61, with  
running times of $O(m\log m)$ and $O(n^3)$ respectively, where 
$m$ and $n$ denote the total number of edges and vertices in the underlying
complete bipartite graph between cities and facilities. In other words,
$m = n_c\times n_f$ and $n=n_c+n_f$, where $n_c$ is the number of 
cities and $n_f$ is the number of facilities.

\subsection{Dual fitting with factor-revealing LP}
The set cover problem offers a particularly simple setting for illustrating
most of the dominant ideas in approximation algorithms (see~\cite{Va.book}).
Perhaps the reason that the method of dual fitting was not clear so far was that the
set cover problem did not require its full power. However, in retrospect, its
salient features are best illustrated again in the simple
setting of the set cover problem -- we do this in Section \ref{disc}.
 
The method of dual fitting can be described as follows, assuming a minimization problem:
The basic algorithm is combinatorial -- in the case of set cover it is in fact
a simple greedy algorithm. Using the linear
programming relaxation of the problem and its dual, one first interprets the
combinatorial algorithm as a primal-dual-type algorithm -- an algorithm
that is iteratively making primal and dual updates. Strictly speaking, this
is not a primal-dual algorithm, since the dual solution computed is, in general,
infeasible (see Section \ref{disc} for a discussion on this issue). 
However, one shows that the primal integral solution found by the 
algorithm is fully paid for by the dual computed.
By {\em fully paid for} we mean that the objective function value of the 
primal solution is bounded by that of the dual. The main step in the
analysis consists of dividing the dual by a suitable factor, say $\gamma$, and showing that
the shrunk dual is feasible, i.e., it {\em fits} into the given instance. The shrunk
dual is then a lower bound on OPT, and $\gamma$ is the approximation
guarantee of the algorithm.

Clearly, we need to find the minimum $\gamma$ that suffices. 
Equivalently, this amounts to finding the worst possible instance --
one in which the dual solution needs to be shrunk the most in order to be
rendered feasible. For each value of $n_c$, the number of cities,
we define a factor-revealing LP that encodes the problem of finding the 
worst possible instance with $n_c$ cities as a linear program. 
This gives a family of LP's, one for
each value of $n_c$. The supremum of the optimal solutions to these LP's is
then the best value for $\gamma$. In our case, we do not know
how to compute this supremum directly. Instead, we obtain a feasible
solution to the dual of each of these LP's. An upper bound on the objective function 
values of these duals can be computed, and is an upper bound on the optimal $\gamma$.
In our case, this upper bound is 1.861 for the first algorithm and 1.61 
for the second one. In order to get a closely matching
tight example, we numerically solve the factor-revealing LP for a large value of $n_c$.

The technique of factor-revealing LPs is similar to the idea of LP bounds in coding theory. 
LP bounds give the best known bounds on the minimum distance of a code with a given 
rate by bounding the solution of a linear program. (cf. McEliece et al.~\cite{MRRW}). 
In the context of approximation algorithms, Goemans and Kleinberg~\cite{mlp} use a 
similar method in the analysis of their algorithm for the minimum latency problem.

\subsection{The facility location problem}

In the (uncapacitated) facility location problem, we have a 
set $\F$ of $n_f$ {\em facilities}
and a set $\C$ of $n_c$ {\em cities}. For every facility $i\in \F$, a nonnegative
number $f_i$ is given as the {\em opening cost} of facility $i$. Furthermore,
for every facility $i\in \F$ and city $j\in \C$, we have a {\em connection
cost} (a.k.a. service cost) $c_{ij}$ between facility $i$ and city $j$.
The objective is to open a subset of the facilities in $\F$, and connect
each city to an open facility so that the total cost is minimized.
We will consider the {\em metric} version of this problem, i.e., the connection costs
satisfy the triangle inequality.

This problem has occupied a central place in operations 
research since the early
60's~\cite{LP, KEH, KH, S1, S2}, and has been studied from the perspectives of
worst case analysis, probabilistic analysis, polyhedral combinatorics and
empirical heuristics (see~\cite{CNW, NW}).
Although the first approximation algorithm for this problem, a greedy
algorithm achieving a guarantee of $O(\log n)$ in the general (non-metric) case
due to Hochbaum~\cite{hochbaum},
dates back to almost 20 years ago, renewed interest in recent years has 
resulted in much progress. 
Recently, the problem has found several new applications
in network design problems such as placement of routers and 
caches~\cite{net1,net2},
agglomeration of traffic or data~\cite{net4,net3}, and web server replications
in a content distribution network (CDN)~\cite{net6,net5,replica}.

The first constant factor approximation algorithm for this problem was given by Shmoys,
Tardos, and Aardal~\cite{STA}. Later, the factor was improved by Chudak 
and Shmoys~\cite{CS} to $1 + 2/e$. Both these 
algorithms were based on LP-rounding, and therefore had high running times.

Jain and Vazirani~\cite{JV} gave a
primal--dual algorithm, achieving a factor of 3, and having the
same running time as ours (we will refer to this as the JV algorithm). Their
algorithm was adapted for solving several related problems such as
the fault-tolerant and outlier versions, and the $k$-median problem
\cite{JV,JVfault,CKMN}.
Mettu and Plaxton~\cite{online} used a restatement of the JV algorithm for
the on-line median problem.

Strategies based on local search and greedy improvement for facility
location problem have also been studied. The work of Korupolu et~al.~\cite{KPR} shows that a simple local search heuristic proposed by Kuehn
and Hamburger~\cite{KH} yields a $(5 + \epsilon)$-approximation 
algorithm with a running time of $O(n^6 \log n /\epsilon)$,
for any $\epsilon > 0$.
Charikar and Guha~\cite{CG} improved the factor slightly to 1.728 by
combining the JV algorithm, greedy augmentation, and the LP-based algorithm~\cite{CS}.
They also combined greedy improvement and cost scaling
to improve the factor of the JV algorithm to 1.853.
For a metric defined by a sparse graph, Thorup~\cite{thorup} has obtained a
$(3+o(1))$-approximation algorithm with running time $\tilde{O}(|E|)$.
Regarding hardness results, Guha and
Khuller~\cite{GK} showed that the best approximation factor possible
for this problem is 1.463, assuming  $NP \not\subseteq DTIME[n^{O(\log \log n)}]$.

Since the publication of the first draft of the present paper, two new algorithms
have been proposed for the facility location problem. The first algorithm, due
to Sviridenko~\cite{1.58}, uses the LP-rounding method to achieve an approximation
factor of 1.58. The second algorithm, due to Mahdian, Ye, and Zhang~\cite{1.52},
combines our second algorithm with the idea of cost scaling to achieve an approximation
factor of 1.52, which is currently the best known factor for this problem.

\subsection{Our results}

Our first algorithm is quite similar to the greedy set cover algorithm:
iteratively pick the most cost-effective choice at each step, where
cost-effectiveness is measured as the ratio of the cost incurred to the
number of new cities served. In order to use LP-duality to analyze this algorithm, we give an alternative 
description which can be seen as a modification of the JV algorithm -- when
a city gets connected to an open facility, it withdraws whatever it has
contributed towards the opening cost of other facilities.
This step of withdrawing contribution is important, since it ensures that
the primal solution is fully paid for by the dual.

The second algorithm has a minor difference with the first one: A city 
might change the facility to which it is connected and connect to 
a closer facility. If so, it offers this difference toward opening the 
latter facility. 

The approximation factor of the algorithms are 1.861 and 
1.61, with running times of $O(m\log m)$ and $O(n^3)$ respectively where
$n$ is the total number of vertices and $m$ is the number of edges 
in the underlying
complete bipartite graph between cities and facilities.

We have experimented our algorithms on randomly generated instances as
well as instances obtained from the Operations Research library~\cite{realdata}
and GT-ITM Internet topology generator~\cite{zegura}.
The cost of the integral solution found is compared against the
solution of the LP-relaxation of the problem, rather than $\OPT$ (computing which 
would be prohibitively time consuming). The results are encouraging: 
The average error of our algorithms is about $3\%$ and $1\%$ respectively, 
and is a significant improvement over the JV algorithm which has an
error of even $100\%$ in some cases. 

The primal-dual algorithm of Jain and Vazirani~\cite{JV} is versatile 
in that it can be used to obtain algorithms for many variants of the 
facility location problem, such as $k$-median~\cite{JV}, a common 
generalization of $k$-median and facility location~\cite{JV}, capacitated
facility location with soft capacities~\cite{JV}, prize collecting 
facility location~\cite{CKMN}, and facility location with 
outliers~\cite{CKMN}. In Section \ref{variants}, we apply our algorithms
to several variants of the problem. First, we consider a common generalization 
of the facility location and $k$-median problems.
In this problem, which we refer to as the {\em $k$-facility location problem},
an instance of the facility location problem and an integer $k$ are given 
and the objective is to find the cheapest solution that opens at most $k$ facilities. 
The $k$-median problem is a special case of this problem in which all opening 
costs are 0. The $k$-median problem is studied 
extensively~\cite{kmed3,CG,CGTS,JV} and the best known approximation algorithm for this problem, 
due to Arya et al.~\cite{kmed3}, achieves a factor of $3+\epsilon$. The $k$-facility 
location problem 
has also been studied in operations research~\cite{CNW}, and the best previously known 
approximation factor for this problem was 6~\cite{JV}.

Next, we show an application of our algorithm to the facility 
location game. We also use our algorithm to improve recent results for some 
other variants of the problem. In the facility location problem with outliers we 
are not required to connect all cities to open facilities. We consider two
versions of this variant: In the robust version, we are allowed to leave $l$ cities 
unconnected. In facility location with penalties we can either 
connect a city to a facility, or pay a specified penalty. Both versions were motivated 
by commercial applications, and were proposed by Charikar~et al.~\cite{CKMN}. 
In this paper we will modify our algorithm to obtain a factor 2 
approximation 
algorithm for these versions, improving the best known result of factor
3~\cite{CKMN}.

In the fault tolerant variant, each city has a specified number of facilities
it should be connected to. This problem was proposed 
in~\cite{JVfault} and the best factor known is 2.47~\cite{GMM}.
We can achieve a factor of 1.61 when all cities have the same 
connectivity requirement.
In addition, we introduce a new variant which can be seen as a special
case of the concave cost version of this problem: the cost of opening
a facility at a location is specified and it can serve exactly one
city. In addition, a {\em setup cost} is charged
the very first time a facility is opened at a given location.

\section{Algorithm 1}
In the following algorithm we use a notion of cost effectiveness. 
Let us say that a
{\em star} consists of one facility and several cities.
The cost of a star is the sum of the opening cost of the facility and the
connection costs between the facility and all the cities in the star.
More formally, the cost of the star $(i,C')$, where
$i$ is a facility and $C' \subseteq C$ is a subset of cities,
is $f_i + \sum_{j \in C'} c_{ij}$. The cost effectiveness
of the star $(i,C')$ is the ratio of the cost of the star
to the size of $C'$, i.e., $({f_i + \sum_{j \in C'} c_{ij}})\left/{|C'|}\right.$.
\medskip

\noindent {\bf Algorithm 1}
\begin{enumerate}
\item Let $U$ be the set of unconnected cities. In the beginning, all cities are unconnected 
i.e. $U := C$ and all facilities are unopened. 
\item While $U \neq \emptyset$:
\begin{itemize}
\item Among all stars, find the most cost-effective one, 
$(i, C')$, open facility $i$, if it is not already open, and connect all cities in $C'$ to $i$.
\item Set $f_i := 0$, $U := U \setminus C'$.
\end{itemize}  
\end{enumerate}
Note that a facility can be chosen again after being opened, but its opening cost is counted only 
once since we set $f_i$ to zero after the first 
time the facility is picked by the algorithm. As far as cities are concerned, every city $j$
is removed from $C$, when connected to an open facility, and is not taken into consideration
again. Also, notice that although the number of stars is exponentially large, in each iteration 
the most cost-effective pair can be found in polynomial time. For each facility $i$, we can sort the cities 
in increasing order of their connection cost to $i$. 
It can be easily seen that the most cost-effective star will consist of a facility and a set, 
containing the first $k$ cities in this order, for some $k$.

The idea of cost effectiveness essentially stems from a similar notion in the greedy algorithm for the 
set cover problem. In that algorithm, the cost effectiveness of a set $S$ is defined to be the cost of 
$S$ over the number of uncovered elements in $S$.
In each iteration, the algorithm picks the most cost-effective set until all elements are covered.
The most cost-effective set can be found either by using direct computation, or by using the dual 
program of the linear programming formulation for the problem. The dual program can also be used 
to prove the approximation factor of the algorithm.
Similarly, we will use the LP-formulation of facility location to analyze our algorithm. 
As we will see, the dual formulation of the problem helps us to understand the nature of the 
problem and the greedy algorithm.

The facility location problem can be captured by an
integer program due to Balinski~\cite{LP}.
For the sake of convenience, we give another equivalent formulation for the problem.
Let ${\cal S}$ be the set of all stars. The facility
location problem can be thought of as picking a minimum cost set of
stars such that each city is in at least one star. This
problem can be captured by the following integer program. In this
program, $x_S$ is an indicator variable denoting whether star $S$
is picked and $c_S$ denotes the cost of star $S$.

%\vspace{-2mm}

\begin{lp}
\label{IP} \minimize & \sum_{S\in {\cal S}} c_S x_S \\[\lpskip]
\st & \forall j\in \C:~ \sum_{S:j \in S} x_S \geq 1 \nonumber
\\
      & \forall S\in {\cal S}:~ x_S \in \{0,1\}  \nonumber
\end{lp}

%\vspace{-2mm}

\noindent The LP-relaxation of this program is:
\begin{lp}
\label{primal} \minimize & \sum_{S\in {\cal S}} c_S x_S
\\[\lpskip] \st & \forall j\in \C:~ \sum_{S:j \in S} x_S \geq 1
\nonumber
\\
      & \forall S\in {\cal S}:~ x_S \geq 0  \nonumber
\end{lp}
%

%\vspace{-2mm}

\noindent The dual program is:
\begin{lp}
\label{dual} \maximize & \sum_{j \in \C} \alpha_{j}\\[\lpskip] \st
& \forall S\in {\cal S}:~  \sum_{j\in S\cap \C}\alpha_j \leq c_S
\nonumber  \\
      & \forall j\in \C:~ \alpha_j \geq 0  \nonumber
\end{lp}

%\vspace{-2mm}

There is an intuitive way of interpreting the dual variables. 
We can think of $\alpha_j$ as the contribution of city $j$, or its share 
toward the total expenses. 
Note that the first inequality of the dual can also be written  as
$\sum_{j \in C}\max(0, \alpha_j-c_{ij}) \leq f_i$ for every facility $i$.
We can now see how the dual variables can help us find the most 
cost-effective star in each iteration of the greedy algorithm: 
if we start raising the dual variables of all unconnected cities 
simultaneously, the most cost-effective star will be the 
first star $(i, C')$ for which $\sum_{j \in C'}\max(0, \alpha_j-c_{ij}) = f_i.$
Hence we can restate Algorithm 1 based on the above observation. 
This is in complete analogy to the greedy algorithm and its restatement 
using LP-formulation for set-cover. \\

\noindent{\bf Restatement of Algorithm 1}                                                                         
\begin{enumerate}
\item We introduce a notion of time, so that each event can be associated with the time 
at which it happened. The algorithm starts at time 0. Initially, each city is defined to 
be unconnected ($U := C$), all facilities are unopened, and $\alpha_j$ is set to 0 for every j.
\item While $U \neq \emptyset$, increase the time, and simultaneously,  for every city $j \in U$,
increase the parameter $\alpha_j$ at the same rate, until one of
the following events occurs (if two events occur at the same
time, we process them in arbitrary order).
\begin{enumerate}
\item For some unconnected city $j$, and some open facility
$i$, $\alpha_j=c_{ij}$. In this case, connect city
$j$ to facility $i$ and remove $j$ from $U$.
\item For some unopened facility $i$, we have $\sum_{j \in U}\max(0, \alpha_j-c_{ij})=f_i.$
This means that the total contribution of the cities is sufficient
to open facility $i$. In this case, open this facility, 
and for every unconnected city $j$
with $\alpha_j\geq c_{ij}$, connect $j$ to $i$, and remove it from $U$.
\end{enumerate}
\end{enumerate}
In each iteration of algorithm 1 the process of opening a facility and/or 
connecting some cities will be defined as an {\em event}. It is easy
to prove the following lemma by induction.
\begin{lemma}
The sequence of events executed by Algorithm 1 and its restatement are identical.
\end{lemma}
\begin{proof}
By induction.
\end{proof}

This restatement can also be seen as a modification of JV algorithm
\cite{JV}. The only difference is that in JV algorithm cities, 
when connected to an open facility, are not excluded from $U$, 
hence they might contribute towards opening several facilities. 
Due to this fact they have a second cleanup phase in which some 
of the already open facilities will be closed down. 

Also, it is worth noting that despite the similarity between Algorithm 1 and 
Hochbaum's greedy algorithm for facility location (which is equivalent to the
set cover algorithm applied on the set of stars), they are not equivalent. This
is because we set $f_i$ to zero after picking a set containing $f_i$. As the following
example shows, the approximation factor of Hochbaum's algorithm is  
$\Omega(\frac{\log n}{\log\log n})$ on instances with metric inequality:
Consider $k$ facilities with opening cost $p^k$ located in the same place
Also $k-1$ groups of cities
$S_1, S_2, \ldots, S_{k-1}$. The group $S_i$ consists of
$p^{k-i+1}$ cities with distance $\sum_{j=1 \ldots i}{p^{j - 1}}$
from the facilities. Other distances are obtained from the triangle
inequality.
Hochbaum's algorithm opens all facilities and therefore its solution 
costs more than $kp^k$. The optimum solution is  $ p^k +
\sum_{i=1 \ldots k-1}\sum_{j=1 \ldots i}{p^{j-1}}$. 
It is easy to show that with a careful choice of $k$, the ratio
of these two expressions is $\Omega(\frac{\log n}{\log\log n})$.
We do not know whether the approximation factor of Hochbaum's algorithm 
on metric instances is strictly less than $\log n$ or not.

\section{Analysis of Algorithm 1}
\label{analysis}
In this section we will give an LP-based analysis of the algorithm.
As stated before, the contribution of each city goes towards opening at
most one facility and connecting the city to an open facility. Therefore, the
total cost of the solution produced by our algorithm will be equal to the 
sum $\sum_j \alpha_j$ of the contributions. However, $\A$ is not a feasible
dual solution as it was in JV algorithm. The reason is that in every
iteration of the restatement of Algorithm 1, we exclude a subset of cities and withdraw 
their contribution from all facilities. So at the end, for some facility
$i$, 
$\sum_j \max(\alpha_j - c_{ij}, 0)$ can be greater 
than $f_i$
and hence the corresponding constraints of the dual program is violated.

However, if we find an $\gamma$ for which $\A/\gamma$ is feasible, 
$\sum_j \alpha_j/\gamma$ would be 
a lower bound to the optimum  and therefore 
the approximation factor of 
the algorithm would be at most $\gamma$. This observation motivates the
following definition.

\noindent{\bf Definition} 
Given $\alpha_j$ ($j = 1, \ldots, n_c$), a facility $i$ is called at most
$\gamma$-overtight if and only if  $$\sum_j \max(\alpha_j/\gamma - c_{ij}, 0) \leq f_i.$$
Using the above definition, it is trivial that 
$\A/\gamma$ is a feasible dual if and only if each facility is at most
$\gamma$-overtight. Now, we want to find such an $\gamma$. Note that 
in the above sum we only need to consider the cities $j$ for which 
$\alpha_j \geq \gamma c_{ij}$. Let us assume without loss of generality that 
it is the case only for the first $k$ cities. Moreover, assume without
loss of generality that $\alpha_1 \leq \alpha_2\leq\cdots  \leq \alpha_{k}$. 
The next two lemmas express the constraints on 
$\A$ imposed by the problem or our algorithm. The first lemma mainly 
captures metric property and the second one expresses the fact that 
the total contribution offered to a facility at any time during the 
algorithm is no more than its cost. 
\begin{lemma}
\label{metricl}
For every two cities $j, j'$ and facility $i$, 
$\alpha_j \leq \alpha_{j'} + c_{ij'} + c_{ij}$.
\end{lemma}
\begin{proof}
If $\alpha_{j'} \geq \alpha_j$, the inequality obviously holds.
Assume $\alpha_j > \alpha_{j'}$.
Let $i'$ be the facility that city $j'$ is connected to by our algorithm. 
Thus, facility $i'$ is open at time $\alpha_{j'}$.  
The contribution $\alpha_j$ cannot be greater than $c_{i'j}$ because in that case 
city $j$ could be connected to facility $i'$ at some time $t < \alpha_j$.
Hence $\alpha_j \leq c_{i'j}$. Furthermore, by triangle inequality,
$c_{i'j} \leq c_{i'j'} + c_{ij'} + c_{ij} \leq  \alpha_{j'} + c_{ij'} + c_{ij}$.
\end{proof}

\begin{lemma} 
\label{contr}
For every city $j$ and facility $i$, $\sum_{l=j}^{k} \max(\alpha_j-c_{il},0)\leq f_i$.
\end{lemma}
\begin{proof}
Assume, for the sake of contradiction, that for some $j$ and some $i$ the inequality does
not hold, i.e., $\sum_{k=j}^{n_c} \max(\alpha_j-c_{ik},0) > f_i$.
By the ordering on cities, for $k \geq j$, $\alpha_k \geq \alpha_j$.
Let time $t = \alpha_j$. 
By the assumption, facility $i$ is fully paid for before time $t$.
For any city $k$, $j \leq k \leq n_c$ for which $\alpha_j - c_{ik} > 0$
the edge $(i, k)$ must be tight before time $t$. 
Moreover, there must be at least one such city. For this city, $\alpha_k < \alpha_j$,
since the algorithm will stop growing $\alpha_k$ as soon as $k$ has a tight
edge to a fully paid for facility. The contradiction establishes the lemma.
\end{proof}

Subject to the constraints introduced by Lemmas \ref{metricl} and
\ref{contr}, we want to find the minimum $\gamma$ for which $\sum_{j=1}^{k}
(\alpha_j/\gamma - c_{ij}) \leq f_i$. In other words, we want to find the 
maximum of the ratio $\frac{\sum_{j=1}^k\alpha_j}{f+\sum_{j=1}^kd_j}$. 
We can define variables $f$, $d_j$, and $\alpha_j$, 
corresponding to facility cost, distances, and 
contributions respectively and write the following maximization program:
\begin{equation}
\label{lp2}
\begin{array}{lll}
z_{k}=&{\rm maximize}& \displaystyle\frac{\sum_{j=1}^k\alpha_j}{f+\sum_{j=1}^kd_j}\\
&{\rm subject\ to\quad}&\begin{array}[t]{lll}
\alpha_j\le\alpha_{j+1} \qquad\qquad\qquad\qquad& \forall j\in\{1,\ldots,k-1\}\\ 
\alpha_j\le\alpha_l+d_j+d_l 		& \forall j,l\in\{1,\ldots,k\}\\ 
\sum_{l=j}^k \max(\alpha_j-d_l,0)\le f 	& \forall j\in\{1,\ldots,k\}\\ 
\alpha_j, d_j, f\ge0			& \forall j\in\{1,\ldots,k\}
\end{array}\end{array}
\end{equation}
It's not difficult to prove that $z_k$ (the maximum value of the objective
function of program~\ref{lp2}) is equal to the optimal solution of the 
following linear program which we call the {\em factor-revealing LP}.
\begin{equation}
\label{LP}
\begin{array}{lll}
z_{k}=&{\rm maximize}& \displaystyle{\sum_{j=1}^k\alpha_j}\\
&{\rm subject\ to\quad}&\begin{array}[t]{lll}
f+\sum_{j=1}^kd_j\le 1\hspace*{2cm}\\
\alpha_j\le\alpha_{j+1} \qquad\qquad\qquad\qquad& \forall j\in\{1,\ldots,k-1\}\\ 
\alpha_j\le\alpha_l+d_j+d_l 		& \forall j,l\in\{1,\ldots,k\}\\ 
x_{jl}\ge\alpha_j-d_l 				& \forall j,l\in\{1,\ldots,k\}\\ 
\sum_{l=j}^k x_{jl}\le f 			& \forall j\in\{1,\ldots,k\}\\ 
\alpha_j, d_j, f\ge0			& \forall j\in\{1,\ldots,k\}
\end{array}\end{array}
\end{equation}
\begin{lemma}
\label{overtight2}
Let $\gamma=\sup_{k\ge1}\{z_k\}$. Every facility is at most $\gamma$-overtight
\end{lemma}
\begin{proof}
Consider facility $i$. We want to show that $\sum_j max(\alpha_j/\gamma - c_{ij},
0) \leq f_i$.  Suppose without loss of generality that  the subset of
cities $j$ such that  $\alpha_j \geq \gamma c_{ij}$ is   $\{j=1, 2, \ldots, k\}$ 
for some $k$. Moreover $\alpha_1\leq\alpha_2\leq\ldots\alpha_{k}$.
Let $d_j = c_{ij}$, $j=1, \ldots, k$, and $f=f_i$. 
By Lemmas~\ref{metricl} and~\ref{contr} it follows immediately
that the constraints of program~\ref{lp2} are satisfied.
Therefore, $\alpha_i, d_i, f$ constitute a feasible solution of
program~\ref{lp2}. Consequently
$ \frac{\sum_{j=1}^k\alpha_j}{f_i+\sum_{j=1}^kc_{ij}}\le z_k.$
\end{proof}

By what we said so far, we know that the approximation factor of 
our algorithm is at most $\sup_{k\ge1}\{z_k\}$. 
In the following theorem, we prove, by demonstrating an infinite family of 
instances, that the approximation ratio of Algorithm 1 is not better than 
$\sup_{k\ge1}\{z_k\}$.

\begin{theorem}
\label{tightex}
The approximation factor of our algorithm is precisely $\sup_{k\ge1}\{z_k\}$.
\end{theorem}
\begin{proof}
Consider an optimum feasible solution of program~\ref{lp2}. We construct
an instance of the facility location problem with $k$ cities and $k+1$
facilities as follows:
The cost of opening
facility $i$ is
$$
f_i=\left\{\begin{array}{ll}0 & {\rm\ if\ }1\le i\le k \\ f & {\rm\ if\ }i=k
+1\end{array}\right.
$$
The connection cost between a city $j$ and a facility $i$ is:
$$
c_{ij}=\left\{\begin{array}{ll}\alpha_j & {\rm\ if\ }1\le i=j\le k \\
d_j & {\rm\ if\ }1\le j\le k, i=k+1 \\
d_i+d_j+\alpha_i & {\rm\ otherwise}
\end{array}\right.
$$
It is easy to see that the connection costs satisfy the triangle inequality.
On this instance,
our algorithm connects city $1$ to facility $1$, then it connects
city $2$ to facility $2$, and finally connects city $k$ to
facility $k$. (The inequality $\sum_{l=j}^k \max(\alpha_j-d_l,0)\le f$
guarantees that city $i$ can get connected to facility $i$ before
facility $k+1$). Therefore, the cost of the restatement of Algorithm 1 is equal to
$\sum_{j=1}^k{c_{jj}}+\sum_{i=1}^kf_i=\sum_{j=1}^k\alpha_j=z_k$.

On the other hand, the optimal solution for this instance is to connect
all the cities to facility $k+1$. The cost of this solution is equal to
$\sum_{j=1}^k c_{k+1,j}+f_{k+1}=f+\sum_{j=1}^kd_j\le 1$.

Thus, our algorithm outputs a solution whose cost is at least $z_k$ times
the cost of the optimal solution.
\end{proof}

The only thing that remains is to find an upper bound on
$sup_{k \geq 1} \{ z_k\}$. 
%The idea is to introduce a feasible
%dual solution for the above linear program.
%This is done in the following lemma. The proof is similar to 
%the proof of Lemma \ref{upperboundlemma}, and is omitted here.
%However, the technique that is used in the proof of the following 
%lemma is explained in detail in Section \ref{solvelpsec}.
%
%
%Idea of the proof that z_k \leq 1.861
%
%
%
%
By solving the factor-revealing LP for 
any particular value of $k$, we get a lower bound
on the value of $\gamma$. In order to prove an upper bound on $\gamma$,
we need to present a general solution to the dual of the factor-revealing LP.
Unfortunately, this is not an easy task in general.
(For example, performing a tight asymptotic analysis of the LP bound 
is still an
open question in coding theory).
However, here empirical results can help us: we can solve the dual of the
factor-revealing LP for 
small values of $k$ to get an idea of how the general optimal solution
looks like. Using this, it is usually possible (although sometimes tedious)
to prove a close-to-optimal upper bound on the value of $z_k$.
We have used this technique to prove an upper bound of $1.861$ on
$\gamma$. 

%
%In order to obtain an upper bound for $sup_{k \geq 1} \{ z_k\}$, 
%we introduce a feasible dual for the above linear program. This dual 
%is not defined explicity. We multiply inequalities in 
%program \ref{lp2} to appropriate coefficients and  
%then we add them together.
 
\begin{lemma}
\label{1861}
For every $k \geq 1$, $z_k \leq 1.861$.
\end{lemma}
\begin{proof}
Let $r=1.8609$. By doubling a feasible solution of \ref{lp2} it is easy 
to show that $z_k \leq z_{2k}$ so we can assume, without loss of generality that $k$
is sufficiently large.
Consider a feasible solution of the program \ref{lp2}.
It is clear from the third inequality that for every $j,j'$ we
have

\begin{equation}
\label{1}
\sum_{i=j}^{j'} (\alpha_j-d_i)\le f.
\end{equation}

Now, we define $l_j$ and $\theta_j$ as follows:

$$
l_j=\left\{\begin{array}{ll}
p_2k & {\rm if\ } j\le p_1k\\
k & j>p_1k
\end{array}\right.$$

$$
\theta_j=\left\{\begin{array}{ll}
\frac{r+1}{p_2k} & {\rm if\ } j\le p_1k\\
\frac{(r+1)(p_2-p_1)}{p_2(1-p_1)k} & p_1k<j\le p_2k\\
0 & j>p_2k
\end{array}\right.$$

where $p_1=0.1991$ and $p_2=0.5696$.
We consider Inequality \ref{1} for every $j\le p_2k$
and $j'=l_j$, and
multiply both sides of this inequality by $\theta_j$.
By adding up all these inequalities, we obtain

\begin{equation}
\label{2}
\sum_{j=1}^{p_1k}\sum_{i=j}^{p_2k}\theta_j(\alpha_j-d_i)
+\sum_{j=p_1k+1}^{p_2k}\sum_{i=j}^{k}\theta_j(\alpha_j-d_i)
\le(\sum_{j=1}^{p_2k}\theta_j)f.
\end{equation}

The coefficient of $f$ in the right-hand side of the above
inequality is equal to
$\sum_{j=1}^{p_2k}\theta_j=
\frac{r+1}{p_2k}p_1k+\frac{(r+1)(p_2-p_1)}{p_2(1-p_1)k}(p_2k-p_1k)
\approx
1.8609<1.861.$
Also, the coefficients of $\alpha_j$ and $d_j$ in the left-hand
side of Inequality \ref{2} are equal to

\begin{equation}
\label{coeffa}
{\rm coeff}[\alpha_j] =
\left\{\begin{array}{ll}
(p_2k-j+1)\theta_j & j\le p_1k \\
(k-j+1)\theta_j & j>p_1k
\end{array}\right.
\end{equation}

\begin{equation}
\label{coeffd}
{\rm coeff}[d_j] =
\left\{\begin{array}{ll}
\sum_{i=1}^j\theta_i & j\le p_2k \\
\sum_{i=p_1k+1}^j\theta_i & j>p_2k
\end{array}\right.
\end{equation}

Notice that the sum of coefficients of $\alpha_j$'s is equal to
\begin{eqnarray*}
\sum_{j=1}^k{\rm coeff}[\alpha_j] & = &
\sum_{j=1}^{p_1k}\frac{r+1}{p_2k}(p_2k-j+1) +
\sum_{j=p_1k+1}^{p_2k}\frac{(r+1)(p_2-p_1)}{p_2(1-p_1)k}(k-j+1) \\
&>&(r+1)\left(p_1-\frac{p_1^2}{2p_2}+
\frac{(p_2-p_1)^2}{p_2(1-p_1)}-\frac{(p_2-p_1)^2(p_1+p_2)}{2p_2(1-p_1)}
\right)k\\
&\approx&1.00004k\\
&>&k
\end{eqnarray*}

Now, we use the inequality $\alpha_i\ge\alpha_j-d_j-d_i$ on
the expression on the left hand side of inequality \ref{2} to
reduce the coefficients of $\alpha_j$'s that are greater than
1, and increase the coefficient of $\alpha_j$'s that are less
than 1. Since the sum of these coefficients is greater than $k$,
using this inequality and the inequality $\alpha_j\ge 0$ we
can obtain an expression $E$ that is less than or equal to the
left hand side of inequality \ref{2}, and in which
all $\alpha_j$'s have coefficient 1. The coefficient of $d_j$
in this expression will be equal to its coefficient in the
left hand side of inequality \ref{2}, plus the absolute value
of the change in the coefficient of the corresponding $\alpha_j$.
Therefore, by equations \ref{coeffa} and \ref{coeffd}
this coefficient is equal to:

$$
{\rm coeff}_E[d_j] =
\left\{\begin{array}{ll}
\sum_{i=1}^j\theta_i + |(p_2k-j+1)\theta_j-1| & j\le p_1k \\
\sum_{i=1}^j\theta_i + |(k-j+1)\theta_j-1| & p_1k<j\le p_2k \\
\sum_{i=p_1k+1}^j\theta_i + |(k-j+1)\theta_j-1| & j>p_2k
\end{array}\right.
$$

If $j\le p_1k$, we have $(p_2k-j+1)\theta_j>
(p_2k-p_1k)\frac{r+1}{p_2k}=(r+1)(p_2-p_1)/p_2\approx 1.8609 > 1$
Therefore,

\begin{eqnarray*}
{\rm coeff}_E[d_j]&=&
\sum_{i=1}^j\theta_i + (p_2k-j+1)\theta_j-1\\
&=&r+O(\frac1k)\\
&<&1.861
\end{eqnarray*}

Similarly, if $p_1k<j\le p_2k$, we have $(k-j+1)\theta_j>
(k-p_2k)\frac{(r+1)(p_2-p_1)}{p_2(1-p_1)k}
=\frac{(r+1)(p_2-p_1)(1-p_2)}{p_2(1-p_1)}\approx 1.00003 > 1$.
Therefore,

\begin{eqnarray*}
{\rm coeff}_E[d_j]&=&
\sum_{i=1}^j\theta_i + (k-j+1)\theta_j-1\\
&=&
r+O(\frac1k)\\
&<& 1.861
\end{eqnarray*}
  
Finally, if $j>p_2k$, the coefficient of $d_j$ is equal to

\begin{eqnarray*}
{\rm coeff}_E[d_j]&=&
\sum_{i=p_1k}^j\theta_i + |0-1|\\
&=&
\frac{(r+1)(p_2-p_1)}{p_2(1-p_1)k}(p_2k-p_1k)+1\\
&\approx&
1.8609\\
&<& 1.861
\end{eqnarray*}

Therefore, in each case, the coefficient of $d_j$ is less than
or equal to $1.861$. Thus, we have proved that

$$\sum_{j=1}^k \alpha_j - \sum_{j=1}^k 1.861d_j<1.861f.$$

This clearly implies that $z_k<1.861$.
\end{proof}

\begin{figure}
\begin{center}
\includegraphics{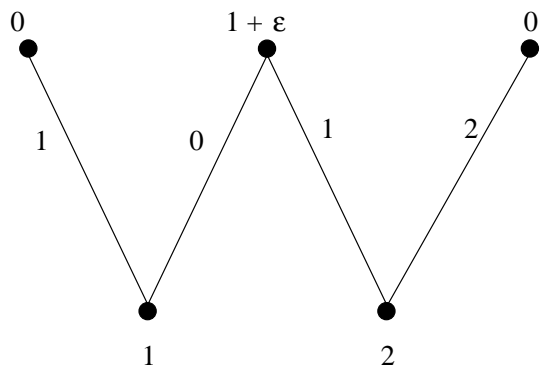}
\end{center}
\caption{The approximation ratio of Algorithm 1 is at least $1.5$}
\label{examplefig}
\end{figure}

Figure \ref{examplefig} shows a tight example for $k=2$, for which
the approximation factor of the algorithm is 1.5. 
The cost of the missing edges is given by triangle inequality.
Numerical computations using the software CPLEX show that 
$z_{300}\approx 1.81$. Thus, the approximation factor of our algorithm 
is between 1.81 and 1.861. We do not know the exact approximation ratio.

\section{Algorithm 2}
\label{algorithm2}
Algorithm 2 is similar to the restatement of Algorithm 1. 
The only difference is that in Algorithm 1 cities stop offering money 
to facilities as soon as they get connected to a facility, but here
they still offer some money to other facilities. 
The amount that an already-connected city
offers to a facility $j$ is equal to the amount that it would save in 
connection cost by switching its
facility to $j$. As we will see in the next section, this change
reduces the approximation factor of the algorithm from 1.861 to 1.61.

\noindent{\bf Algorithm 2}                                                                         
\begin{enumerate}
\item We introduce a notion of time. The algorithm starts at time 0. At this time, 
each city is defined to 
be unconnected ($U := C$), all facilities are unopened, and $\alpha_j$ 
is set to 0 for every j.

At every 
moment, each city $j$ offers some money from its contribution
to each {\em unopened} facility $i$. The amount of this 
offer is computed as follows: If $j$ is unconnected,
the offer is equal to $\max(\alpha_j-c_{ij},0)$ (i.e., if 
the contribution of $j$ is more than the cost that it has
to pay to get connected to $i$, it offers to pay this 
extra amount to $i$); If $j$ is already connected to
some other facility $i'$, then its offer to facility 
$i$ is equal to $\max(c_{i'j}-c_{ij},0)$ (i.e., the
amount that $j$ offers to pay to $i$ is equal to the 
amount $j$ would save by switching its facility from $i'$ to $i$).
\item While $U \neq \emptyset$, increase the time, and simultaneously,  for every city $j \in U$,
increase the parameter $\alpha_j$ at the same rate, until one of
the following events occurs (if two events occur at the same
time, we process them in an arbitrary order).
\begin{enumerate}
\item For some unopened facility $i$, the total offer that it receives from cities is
equal to the cost of opening $i$. In this case, we open facility $i$,
and for every city $j$ (connected or unconnected) which has a non-zero offer to $i$,
we connect $j$ to $i$. The amount that $j$ had offered to $i$ is now called the {\em
contribution} of $j$ toward $i$, and $j$ is no longer allowed to decrease this contribution.
\item For some unconnected city $j$, and some open facility
$i$, $\alpha_j=c_{ij}$. In this case, connect city
$j$ to facility $i$ and remove $j$ from $U$.
\end{enumerate}
\end{enumerate}
Clearly the main issue in the facility location problem is to decide which
facilities to open. Once this is done, each city should be connected to
the closest open facility. Observe that Algorithm 2 makes greedy choices
in deciding which facilities to open and once it opens a facility, it does not
alter this decision. In this sense, it is also a greedy algorithm.

\section{Analysis of Algorithm 2}
\label{analysis2}
The following fact should be obvious from the description of Algorithm 2.
\begin{lemma}
\label{cost}
The total cost of the solution found by Algorithm 2 is equal to the sum
of $\alpha_j$'s.
\end{lemma}
Now, as in the analysis of Algorithm 1, we need to find a number $\gamma$, such
that for every star $S$, $\sum_{j\in S\cap \C}\alpha_j \leq \gamma c_S$.
Such a $\gamma$ will be an upper bound on the approximation ratio of the
algorithm, since if for every facility $i$ that is opened in the optimal solution
and the collection $A$ of cities that are connected to it, we write the inequality
$\sum_{j\in A}\alpha_j \leq \gamma (f_i + \sum_{j\in A}c_{ij})$ and add up these
inequalities, we will obtain that the cost of our solution is at most $\gamma$ times
the cost of the optimal solution.

\subsection{Deriving the factor-revealing LP}
\label{factorlpsec}
Our proof follows the methodology of Section \ref{analysis}:
express various constraints that are imposed by the problem or
by the structure of the algorithm as inequalities and get a bound on the
value of $\gamma$ defined above by solving a series of linear programs.

Consider a star $S$ consisting of a facility having opening cost $f$ (with a slight
misuse of the notation, we call this facility $f$), and $k$ cities numbered
1 through $k$. Let $d_j$ denote the connection cost between facility $f$ and city $j$,
and $\alpha_j$ denote the contribution of the city $j$ at the end of Algorithm 2.
We may assume without loss of generality that

\begin{equation}
\label{ineq1}
\alpha_1\le\alpha_2\le\cdots\le\alpha_k.
\end{equation}

We need more variables to capture the execution of Algorithm 2.
For every $i$ ($1\le i\le k$), consider the situation of the 
algorithm at time $t=\alpha_i-\ep$,
where $\ep$ is very small, i.e., just a moment before city $i$ 
gets connected for the first time.
At this time, each of the cities $1,2,\ldots,i-1$ might be 
connected to a facility. For every $j<i$,
if city $j$ is connected to some facility at time $t$, 
let $r_{j,i}$ denote the connection cost
between this facility and city $j$; otherwise, let $r_{j,i}:=\alpha_j$. The latter
case occurs if and only if $\alpha_i=\alpha_j$. It turns out that these variables
($f$, $d_j$'s, $\alpha_j$'s, and $r_{j,i}$'s) are enough to write down some
inequalities to bound the ratio of the sum of $\alpha_j$'s to the cost of
$S$ (i.e., $f+\sum_{j=1}^kd_j$).

First, notice that once a city gets connected to a
facility, its contribution remains constant and it cannot revoke its contribution
to a facility, so it can never get connected to another facility with a higher
connection cost. This implies that for every $j$,

\begin{equation}
\label{ineq2}
r_{j,j+1}\ge r_{j,j+2}\ge\cdots\ge r_{j,k}.
\end{equation}

Now, consider time $t=\alpha_i-\ep$. At this time, the amount 
city $j$ offers to facility $f$ is equal to

%\begin{equation}
\[
%\displaystyle\left\{
\begin{array}{ll}
\max(r_{j,i}-d_j,0) & {\rm\ if\ }j<i, {\rm\ and}\\
\max(t-d_j,0) & {\rm\ if\ }j\ge i.
\end{array}\hspace*{8cm}
%\right.
\]
%\end{equation}

Notice that by the definition of $r_{j,i}$ this holds even if $j<i$ 
and $\alpha_i=\alpha_j$.
It is clear from Algorithm 2 that the total offer of cities to a facility
can never become larger than the opening cost of the facility.
Therefore, for all $i$,

%\mathindent 1.5em
\begin{equation}
\label{ineq3}
\sum_{j=1}^{i-1}\max( r_{j,i}-d_j,0)+\sum_{j=i}^k \max(\alpha_i-d_j,0)\le f.
\end{equation}
\mathindent 3em

The triangle inequality is another important constraint that we need to use.
Consider cities $i$ and $j$ with $j<i$ at time $t=\alpha_i-\epsilon$.  
Let $f'$ be the facility $j$ is connected to at time $t$.
By the triangle inequality and the definition of $r_{j,i}$, the connection cost 
$c_{f'i}$ between city $i$ and facility $f'$ is at most
$r_{j,i}+d_i+d_j$. Furthermore, $c_{f'i}$ can not be less than $t$, since if it 
is, our algorithm could have connected the city $i$ to the facility $f'$ at a 
time earlier than $t$, which is a contradiction. Here we need to be careful with 
the special case $\alpha_i=\alpha_j$. In this case, $r_{j,i}+d_i+d_j$ is not more
than $t$. If $\alpha_i\neq \alpha_j$, the facility $f'$ is open at time $t$ and
therefore city $i$ can get connected to it, if it can pay the connection cost.
Therefore for every $1\le j<i\le k$,

\begin{equation}
\label{ineq4}
\alpha_i\le r_{j,i}+d_i+d_j.
\end{equation}

The above inequalities form the following factor-revealing LP.

\mathindent 0.5em
\begin{lp}
\label{biglp} \maximize & \frac{\sum_{i=1}^k\alpha_i}{f+\sum_{i=1}^kd_i}\\[\lpskip]
\st & \forall\,1\le i<k:~\alpha_i\le\alpha_{i+1}\nonumber \\
      & \forall\,1\le j<i< k:~ r_{j,i}\ge r_{j,i+1}\nonumber \\
      & \forall\,1\le j<i\le k:~ \alpha_i\le r_{j,i}+d_i+d_j \nonumber \\
      & \forall\,1\le i\le k:~ \sum_{j=1}^{i-1}\max( r_{j,i}-d_j,0)
\nonumber\\[-2.5mm]&\hspace{20mm}+\sum_{j=i}^k \max(\alpha_i-d_j,0)\le f \nonumber\\
      & \forall\, 1\le j\le i\le k:~ \alpha_j, d_j, f,  r_{j,i}\ge0  \nonumber
\end{lp}
\mathindent 3em

Notice that although the above optimization program
is not written in the form of a linear program, it is easy to change it to a linear
program by introducing new variables and inequalities.

\begin{lemma}
\label{mainlemma}
If $z_k$ denotes the solution of the factor-revealing LP,
then for every star $S$ consisting of a facility and $k$ cities, the sum
of $\alpha_j$'s of the cities in $S$ in Algorithm 2 is at most $z_kc_S$.
\end{lemma}
\begin{proof}
Inequalities \ref{ineq1}, \ref{ineq2}, \ref{ineq3}, and \ref{ineq4} derived above
imply that the values $\alpha_j,d_j,f,r_{j,i}$
that we get by running Algorithm 2 constitute a feasible solution of the
factor-revealing LP. Thus, the value of the objective function
for this solution is at most $z_k$.
\end{proof}

\noindent Lemmas \ref{cost} and \ref{mainlemma} imply the following.

\begin{lemma}
\label{gammaapprox}
Let $z_k$ be the solution of the factor-revealing LP, and $\gamma:=\sup_k\{z_k\}$.
Then Algorithm 2 solves the metric facility location problem with an approximation
factor of $\gamma$.
\end{lemma}

\subsection{Solving the factor-revealing LP}
\label{solvelpsec}
As mentioned earlier, the optimization program (\ref{biglp}) can be written as a
linear program. This enables
us to use an LP-solver to solve the factor-revealing LP for small values of $k$, in order to
compute the numerical value of $\gamma$. Table \ref{table} shows a summary of results
that are obtained by solving the factor-revealing LP using CPLEX.
It seems from the experimental results that $z_k$ is an increasing sequence that
converges to some number close to $1.6$ and hence $\gamma\approx 1.6$.

\begin{table}
{\small
$$\begin{tabular}{|l|l|} \hline
$k$ & $\max_{i\le k}z_i$ \\\hline
10  &  1.54147\\
20  &  1.57084\\
50  &  1.58839\\
100 &  1.59425\\
200 &  1.59721\\
300 &  1.59819\\
400 &  1.59868\\
500 &  1.59898\\\hline
\end{tabular}$$
}
\caption{Solution of the factor-revealing LP}
\label{table}
\end{table}

We are using the same idea as Lemma \ref{1861} in Section \ref{analysis} 
to prove the upper bound of 1.61 on $z_k$. 

\begin{lemma}
\label{upperboundlemma}
Let $z_k$ be the solution to the factor-revealing LP. Then for every $k$, $z_k\le 1.61$.
\end{lemma}
\begin{proof}
Using the same argument as in Lemma \ref{1861},  we can assume, without loss of generality, that
$k$ is sufficiently large.
Consider a feasible solution of the factor-revealing LP. Let $x_{j, i} := max(r_{j, i} - d_j, 0)$.
The fourth inequality of the factor-revealing LP implies that for every $i\le i'$,

\begin{equation}
\label{eq1}
(i'-i+1)\alpha_i\le \sum_{j=i}^{i'} d_j + f - \sum_{j=1}^{i - 1} x_{j,i}.
\end{equation}

Now, we define $l_i$ as follows:

\[l_i=\left\{\begin{array}{ll}
p_2k & {\rm\ if\ } i\le p_1k\\
k & {\rm\ if\ } i>p_1k
\end{array}\right.\]

where $p_1$ and $p_2$ are two constants (with $p_1<p_2$) that will be fixed later.
Consider Inequality \ref{eq1} for every $i\le p_2k$
and $i'=l_i$, and divide both sides of this inequality by $(l_i-i+1)$.
By adding up these inequalities we obtain

\begin{eqnarray}
\label{eq2}
\sum_{i=1}^{p_2k}\alpha_i&\le&
\sum_{i=1}^{p_2k}\sum_{j=i}^{l_i}\frac{d_j}{l_i-i+1}
+(\sum_{i=1}^{p_2k}\frac1{l_i-i+1})f%\nonumber\\&&
-\sum_{i=1}^{p_2k}\sum_{j=1}^{i - 1}\frac{x_{j,i}}{l_i-i+1}.
\end{eqnarray}

Now for every $j\le p_2k$, let $y_j := x_{j, p_2k}$. The second inequality of the factor-revealing LP implies
that $x_{j,i}\ge y_j$ for every $j<i\le p_2k$ and $x_{j,i}\le y_j$ for every $i>p_2k$.
Also, let $\zeta:=\sum_{i=1}^{p_2k}\frac1{l_i-i+1}$.
Therefore, inequality \ref{eq2} implies

\mathindent 1em
\begin{eqnarray}
\label{eq3}
\sum_{i=1}^{p_2k}\alpha_i&\le&
\sum_{i=1}^{p_2k}\sum_{j=i}^{l_i}\frac{d_j}{l_i-i+1}
+\zeta f%\nonumber\\&&
-\sum_{i=1}^{p_2k}\sum_{j=1}^{i - 1}\frac{y_j}{l_i-i+1}.
\end{eqnarray}
\mathindent3em

Consider the index $\ell\le p_2k$ for which $2d_\ell + y_\ell$ has its
minimum (i.e., for every $j\le p_2k$, $2d_\ell + y_\ell\le 2d_j + y_j$).
The third inequality of the factor-revealing LP implies that for $i=p_2k+1,\ldots,k$,

\mathindent 1em
\begin{eqnarray}
\label{eq4}
\alpha_i\le r_{\ell,i}+d_i+d_\ell\le x_{\ell,i}+2d_\ell+d_i\le d_i + 2d_\ell + y_\ell.
\end{eqnarray}
\mathindent 3em

By adding Inequality \ref{eq4} for $i=p_2k+1,\ldots,k$
with Inequality \ref{eq3} we obtain

\mathindent 1em
\begin{eqnarray*}
\sum_{i=1}^{k}\alpha_i &\le&
\sum_{i=1}^{p_2k}\sum_{j=i}^{l_i}\frac{d_j}{l_i-i+1}
+ (2d_\ell + y_\ell)(1-p_2)k %\\&&
+ \sum_{j=p_2k+1}^k d_j-\sum_{i=1}^{p_2k}\sum_{j=1}^{i - 1}\frac{y_j}{l_i-i+1}
+ \zeta f\\
&=&\sum_{j=1}^{p_2k}\zeta d_j
-\sum_{j=1}^{p_2k}\sum_{i=j+1}^{p_2k}\frac{d_j+y_j}{l_i-i+1}%+ (2d_\ell + y_\ell)(1-p_2)k
%\\&&
+ \sum_{j=p_2k+1}^k (1+\sum_{i=p_1k+1}^{p_2k}\frac1{k-i+1})d_j%+ \zeta f
\\&&+ (2d_\ell + y_\ell)(1-p_2)k + \zeta f\\
&\le&\sum_{j=1}^{p_2k}\zeta d_j+ \sum_{j=p_2k+1}^k (1+\sum_{i=p_1k+1}^{p_2k}\frac1{k-i+1})d_j + \zeta f
\\&& + (2d_\ell + y_\ell)\left((1-p_2)k-\frac12\sum_{j=1}^{p_2k}\sum_{i=j+1}^{p_2k}\frac1{l_i-i+1}\right),
\end{eqnarray*}
\mathindent 3em
where the last inequality is a consequence of the inequality $2d_\ell + y_\ell\le 2d_j + y_j \le 2d_j+2y_j$
for $j\le p_2k$. Now, let $\zeta':=1+\sum_{i=p_1k+1}^{p_2k}\frac1{k-i+1}$ and
$\delta:=(1-p_2)-\frac1{2k}\sum_{j=1}^{p_2k}\sum_{i=j+1}^{p_2k}\frac1{l_i-i+1}$. Therefore,
the above inequality can be written as follows:

\mathindent 1em
\begin{eqnarray}
\label{maineq}
\sum_{i=1}^{k}\alpha_i&\le&
\sum_{j=1}^{p_2k}\zeta d_j + \sum_{j=p_2k+1}^k\zeta'd_j%\nonumber\\&&
 + \zeta f+\delta(2d_\ell + y_\ell)k,
\end{eqnarray}
\mathindent 3em
where
\mathindent 1em
\begin{eqnarray}
\label{eqzeta}
\zeta&=&\sum_{i=1}^{p_2k}\frac1{l_i-i+1}%\nonumber\\
%&=&\sum_{i=1}^{p_1k}\frac1{p_2k-i+1}+\sum_{i=p_1k+1}^{p_2k}\frac1{k-i+1}\nonumber\\
%&=&H_{p_2k}-H_{p_2k-p_1k}+H_{k-p_1k}-H_{k-p_2k}\nonumber\\
%&=&
=\ln\frac{p_2(1-p_1)}{(p_2-p_1)(1-p_2)}+o(1),\\
\label{eqzetap}
\zeta'&=&1+\sum_{i=p_1k+1}^{p_2k}\frac1{k-i+1}%\nonumber\\
%&=&1+H_{k-p_1k}-H_{k-p_2k}\nonumber\\
%&=&
=1+\ln\frac{1-p_1}{1-p_2}+o(1), %{\rm\ and}
\\
\label{eqdelta}
\delta&=&1-p_2-\frac1{2k}\sum_{j=1}^{p_2k}\sum_{i=j+1}^{p_2k}\frac1{l_i-i+1}\nonumber\\
%&=&1-p_2-\frac1{2k}\sum_{i=1}^{p_2k}\frac{i-1}{l_i-i+1} \nonumber\\
%&=&1-p_2+\frac1{2k}\left(p_2k-p_2k(H_{p_2k}-H_{(p_2-p_1)k})\right.\nonumber\\
%&&\left.-k(H_{(1-p_1)k}-H_{(1-p_2)k})\right)\nonumber\\
&=&\frac12(2-p_2-p_2\ln\frac{p_2}{p_2 - p_1}-\ln\frac{1-p_1}{1-p_2}) + o(1).
\end{eqnarray}
\mathindent 3em

Now if we choose $p_1$ and $p_2$ such that $\delta<0$,
and let $\gamma:=\max(\zeta,\zeta')$ then inequality \ref{maineq}
implies that

\[
\sum_{i=1}^{k}\alpha_i \le (\gamma+o(1))(f+\sum_{i=1}^{k}d_j).
\]

Using equations \ref{eqzeta}, \ref{eqzetap}, and \ref{eqdelta}, it is easy to
see that subject to the condition $\delta<0$, the value of $\gamma$ is
minimized when $p_1\approx0.439$ and $p_2\approx0.695$, which gives us
$\gamma<1.61$.
\end{proof}

Also, as in the proof of Theorem \ref{tightex}, we can use the optimal solution 
of the factor-revealing LP that is computed numerically (see Table \ref{table})
to construct an example on which our algorithm performs at least $z_k$ times
worse than the optimum. These results imply the following. 

\begin{theorem}
\label{mainthm}
Algorithm 2 solves the facility location problem in time $O(n^3)$, where $n=\max(n_f,n_c)$,
with an approximation ratio between 1.598 and 1.61.
\end{theorem}

\section{The tradeoff between facility and connection costs}
\label{tradeoffsec}
We defined the cost of a solution in the facility location problem as the sum
of the facility cost (i.e., total cost of opening facilities) and the connection
cost. We proved in the previous section
that Algorithm 2 achieves an overall performance guarantee of 1.61. However,
sometimes it is useful to get different approximation guarantees for facility
and connection costs. The following theorem gives such a guarantee.
The proof is similar to the proof of Lemma \ref{gammaapprox}.

\begin{theorem}
\label{tradeoffthm}
Let $\gamma_f\ge1$ and $\gamma_c:=\sup_k\{z_k\}$, where $z_k$ is the solution
of the following optimization program.
\mathindent 0.5em
\begin{lp}
\label{tradeofflp} \maximize & \frac{\sum_{i=1}^k\alpha_i - \gamma_f f}{\sum_{i=1}^kd_i}\\[\lpskip]
\st & \forall\,1\le i<k:~\alpha_i\le\alpha_{i+1}\nonumber \\
      & \forall\,1\le j<i< k:~ r_{j,i}\ge r_{j,i+1}\nonumber \\
      & \forall\,1\le j<i\le k:~ \alpha_i\le r_{j,i}+d_i+d_j \nonumber \\
      & \forall\,1\le i\le k:~ \sum_{j=1}^{i-1}\max( r_{j,i}-d_j,0)
\nonumber\\[-2.5mm]&\hspace{21mm}+\sum_{j=i}^k \max(\alpha_i-d_j,0)\le f \nonumber\\
      & \forall\, 1\le j\le i\le k:~ \alpha_j, d_j, f,  r_{j,i}\ge0  \nonumber
\end{lp}
\mathindent 3em
Then for every instance $\cal I$ of the facility location problem, and for every
solution SOL for $\cal I$ with facility cost $F_{SOL}$ and connection cost $C_{SOL}$, the
cost of the solution found by Algorithm 2 is at most $\gamma_fF_{SOL}+\gamma_cC_{SOL}$.
\end{theorem}
\fig{trdffig}{3.5cm}{The tradeoff between $\gamma_f$ and $\gamma_c$}{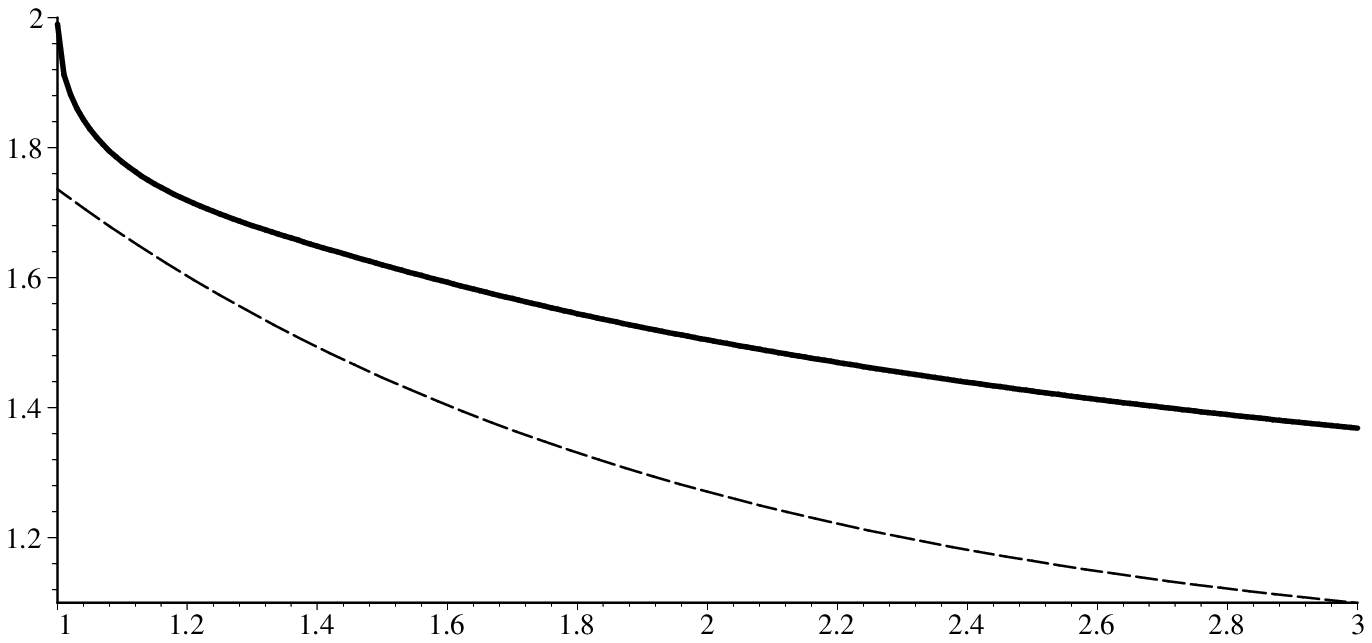}
We have computed the solution of the optimization program \ref{tradeofflp} for $k=100$, and
several values of $\gamma_f$ between 1 and 3, to get an estimate of the corresponding $\gamma_c$'s.
The result is shown in the diagram in Figure \ref{trdffig}. Every point $(\gamma_f,\gamma'_c)$ on
the thick line in this diagram represents a value of $\gamma_f$, and the corresponding
estimate for the value of $\gamma_c$. The dashed line shows the following lower bound,
which can be proved easily by adapting the proof of Guha and Khuller~\cite{GK} for
hardness of the facility location problem.

\begin{theorem}
\label{lmphard}
Let $\gamma_f$ and $\gamma_c$ be constants with $\gamma_c<1+2e^{-\gamma_f}$.
Assume there is an algorithm $\cal A$ such that for every instance $\cal I$ 
of the metric facility location problem, $\cal A$ finds a solution whose cost 
is not more than $\gamma_fF_{SOL}+\gamma_cC_{SOL}$ for every solution $SOL$ 
for $\cal I$ with facility and connection costs $F_{SOL}$ and $C_{SOL}$.
Then $\NP \subseteq {\rm DTIME}[n^{O(\log \log n)}]$.
\end{theorem}

Similar tradeoff problems are considered by Charikar and Guha~\cite{CG}. 
However, an important
advantage that we get here is that all the inequalities $ALG\le \gamma_fF_{SOL}+\gamma_cC_{SOL}$
are satisfied by a {\em single} algorithm.
In Section \ref{variants}, we will use the point $\gamma_f=1$ of this tradeoff
to design algorithms for other variants of the facility location problem.
Other points of this tradeoff can also be useful in designing other algorithms
based on our algorithm. For example, Mahdian, Ye, and Zhang~\cite{1.52} use
the point $\gamma_f=1.1$ of this tradeoff to obtain a 1.52-approximation
algorithm for the metric facility location problem.

\section{Experimental Results}
We have implemented our algorithms, as well as the JV algorithm, using 
the programming language C. We have made four kinds of experiments.
In all cases the solution of the algorithms is compared to 
the optimal solution of the LP-relaxation, computed using the package
CPLEX to obtain an upper bound on the approximation factor of the algorithms.
                                               
The test bed of our first set of experiments consists of
randomly generated instances on a $10,000 \times 10,000$ grid: In each
instance, cities and facilities are points, drawn randomly from the
grid.  The connection cost
between a city and a facility is set to be equal to the euclidean distance of
the corresponding points. Furthermore, the opening cost of each
facility is drawn uniformly at random from the integers  between 0 and 9999. 

For the second set of experiments, we have generated random graphs
(according to the distribution $G(n,p)$) and assigned uniform 
random weights on the edges. Cities and facilities correspond to the
nodes of this graph, and the connection cost between a city and a facility 
is defined to be the shortest path between the corresponding nodes. 
The opening costs of facilities are generated at random.

The instance sizes in both of the above types vary from 50 cities
and 20 facilities to 400 cities and 150 facilities. For each size, 
15 instances are generated and the average error of the algorithm 
(compared to the LP lower bound) is computed. 
The results of these experiments are shown in Table~\ref{grid-rand}.

\begin{table}
\begin{center}
\begin{tabular}{|l|l|l|l|l|l|l|l|} \hline
     &       & \multicolumn{3}{c|}{Random Points on a Grid} & \multicolumn{3}{c|}{Random Graphs} \\
\cline{3-8}
$n_c$ & $n_f$ & JV & ALG 1 & ALG 2 & JV & ALG 1 & ALG 2\\ \hline  \hline
50 & 20 & 1.0927 & 1.0083 & 1.0004 & 1.0021 & 1.0007 & 1.0001 \\ \hline
100 & 20 & 1.0769 & 1.0082 & 1.0004 & 1.0014 & 1.0022 & 1.0 \\ \hline
100 & 50 & 1.2112 & 1.0105 & 1.0013 & 1.0225 & 1.0056 & 1.0005 \\ \hline
200 & 50 & 1.159 & 1.0095 & 1.001 & 1.0106 & 1.0094 & 1.0002 \\ \hline
200 & 100 & 1.301 & 1.0105 & 1.0016 & 1.0753 & 1.0178 & 1.0018 \\ \hline
300 & 50 & 1.1151 & 1.0091 & 1.0011 & 1.0068 & 1.0102 & 1.0002 \\ \hline
300 & 80 & 1.1787 & 1.0116 & 1.001 & 1.0259 & 1.0171 & 1.0004 \\ \hline
300 & 100 & 1.2387 & 1.0118 & 1.0014 & 1.0455 & 1.0185 & 1.0009\\ \hline
300 & 150 & 1.327 & 1.0143 & 1.0015 & 1.1365 & 1.0249 & 1.0018 \\ \hline
400 & 50 & 1.0905 & 1.0092 & 1.0005 & 1.0044 & 1.012 & 1.0 \\ \hline
400 & 100 & 1.8513 & 1.0301 & 1.0026 & 1.0313 & 1.0203 & 1.0003 \\ \hline
400 & 150 & 1.8112 & 1.0299 & 1.0023 & 1.1008 & 1.0234 & 1.0009 \\ \hline
\end{tabular}
\end{center}
\caption{Random Graphs and Random Points on a Grid}
\label{grid-rand}
\end{table}

An Internet topology generator software, namely GT-ITM, is used to generate 
the third set of instances. GT-ITM is a software package for generating graphs 
that have a structure modeling the topology of the Internet~\cite{zegura}. 
This model is used because of the applications of facility location problems
in network applications such as placing web server replicas~\cite{replica}. 
In this model we consider transit nodes as
potential facilities and stub nodes as cities. The connection cost is
the distance produced by the generator.  The opening costs are again
random numbers. We have generated 10 instances for each of the 10 different instance
sizes. The results are shown in Table~\ref{gtitm}.

We also tested all algorithms on 15 instances 
from~\cite{realdata}, which is a library of test data
sets for several operations research problems. Our results are
shown in Table~\ref{or}.

\begin{table}
\begin{center}
\begin{tabular}{|l|l|l|l|l|} \hline
$n_c$ & $n_f$ & JV & ALG 1 & ALG 2 \\ \hline  \hline
100 & 20 & 1.004 & 1.0047 & 1.0001 \\ \hline
160 & 20 & 1.5116 & 1.0612 & 1.0009 \\ \hline
160 & 40 & 1.065 & 1.0063 & 1.0 \\ \hline
208 & 52 & 2.2537 & 1.074 & 1.019 \\ \hline
240 & 60 & 1.0083 & 1.0045 & 1.0001 \\ \hline
300 & 75 & 1.8088 & 1.0478 & 1.0006 \\ \hline
312 & 52 & 1.7593 & 1.0475 & 1.0008 \\ \hline
320 & 32 & 1.0972 & 1.0015 & 1.0 \\ \hline
400 & 100 & 1.0058 & 1.0048 & 1.0 \\ \hline
416 & 52 & 1.0031 & 1.0048 & 1.0 \\ \hline
\end{tabular}
\end{center}
\caption{GT-ITM Model}
\label{gtitm}
\end{table}

\begin{table}
\begin{center}
\begin{tabular}{|l|l|l|l|l|} \hline
$n_c$ & $n_f$ & JV & ALG 1 & ALG 2 \\ \hline  \hline
50 & 16 & 1.0642 & 1.0156 & 1.0 \\ \hline
50 & 16 & 1.127 & 1.0363 & 1.0 \\ \hline
50 & 16 & 1.1968 & 1.0258 & 1.0 \\ \hline
50 & 16 & 1.2649 & 1.0258 & 1.0022 \\ \hline
50 & 25 & 1.1167 & 1.006 & 1.0028 \\ \hline
50 & 25 & 1.2206 & 1.0393 & 1.0 \\ \hline
50 & 25 & 1.3246 & 1.0277 & 1.0 \\ \hline
50 & 25 & 1.4535 & 1.0318 & 1.0049 \\ \hline
50 & 50 & 1.3566 & 1.0101 & 1.0017 \\ \hline
50 & 50 & 1.5762 & 1.0348 & 1.0061 \\ \hline
50 & 50 & 1.7648 & 1.0378 & 1.0022 \\ \hline
50 & 50 & 2.0543 & 1.0494 & 1.0075 \\ \hline
1000 & 100 & 1.0453 & 1.0542 & 1.0023 \\ \hline
1000 & 100 & 1.0155 & 1.0226 & 1.0 \\ \hline
1000 & 100 & 1.0055 & 1.0101 & 1.0 \\ \hline
\end{tabular}
\end{center}
\caption{Instances from Operations Research library}
\label{or}
\end{table}

As we can see from the tables, Algorithm 2 behaves extremely
well, giving almost no error in many cases. Algorithm 1 has an error of 7\%
on the worst instance and an average error of 2-3\%. 
On the other hand, the JV algorithm has much larger error,
sometimes as high as 100 \%. We should also
note that the running times of the three algorithms did not vary
significantly. In the biggest instances of 1000 cities and 100
facilities all the algorithms ran in approximately 1-2 seconds.
The implementation of the algorithms as well as all the data sets are
available upon request. For other experimental results see \cite{low}.

\section{Variants of the problem}
\label{variants}
In this section, we show that our algorithms can also be applied to several 
variants of the metric facility location problem.
\subsection{The $k$-median problem}
\label{kmediansec}
The {\em $k$-median problem} differs from the facility location problem
in two respects: there is no cost for opening facilities, and
there is an upper bound $k$, that is supplied as part of the input,
on the number of facilities that can be opened. 
The {\em $k$-facility location problem} is a common generalization of
$k$-median and the facility location problem. In this problem, we have
an upper bound $k$ on the number of facilities that can be opened, as
well as costs for opening facilities.
The $k$-median problem is studied extensively~\cite{kmed3,CG,CGTS,JV} and 
the best known approximation algorithm for this problem, 
due to Arya et al.~\cite{kmed3}, achieves a factor of $3+\epsilon$. 
It is also straightforward to adapt the proof of hardness of the
facility location problem~\cite{GK} to show that 
there is no $(1+\frac2e-\ep)$-approximation
algorithm for $k$-median, unless $\NP \subseteq {\rm DTIME}[n^{O(\log \log n)}]$.
Notice that this proves that $k$-median is a strictly harder 
problem to approximate than the facility location problem because the latter
can be approximated within a factor of 1.61.

Jain and Vazirani~\cite{JV} reduced the $k$-median problem to the facility
location problem in the following sense:
Suppose ${\cal A}$ is an approximation algorithm for the
facility location problem. Consider an instance $\cal I$ of the
problem with optimum cost $OPT$, and let $F$ and $C$ be the facility
and connection costs of the solution found by $\cal A$.
We call algorithm ${\cal A}$ a Lagrangian Multiplier Preserving
$\alpha$-approximation (or LMP $\alpha$-approximation for short)
if for every instance $\cal I$, $C \leq \alpha(OPT - F).$
Jain and Vazirani~\cite{JV} show that an LMP $\alpha$-approximation
algorithm for the metric facility location problem gives rise to a
$2\alpha$-approximation algorithm for the metric $k$-median problem.
They have noted that this result also holds for the $k$-facility
location problem.

\begin{lemma}{\bf\cite{JV}}
\label{kmedian}
An LMP $\alpha$-approximation algorithm for the facility
location problem gives a $2\alpha$-approximation algorithm for the
$k$-facility location problem.
\end{lemma}

Here we use Theorem
\ref{tradeoffthm} together with the scaling technique of Charikar and Guha~\cite{CG}
to give an LMP $2$-approximation algorithm for the 
metric facility location problem based on Algorithm 2. This will result 
in a $4$-approximation algorithm for the metric $k$-facility location 
problem, whereas the best previously known was 6~\cite{JV}.

\begin{lemma}
\label{scalinglemma}
Assume there is an algorithm $\cal A$ for the metric facility location problem
such that for every instance $\cal I$ and every solution $SOL$ for $\cal I$,
$\cal A$ finds a solution of cost at most $F_{SOL}+\alpha C_{SOL}$,
where $F_{SOL}$ and $C_{SOL}$ are facility and connection costs of $SOL$,
and $\alpha$ is a fixed number. Then there is an LMP $\alpha$-approximation
algorithm for the metric facility location problem.
\end{lemma}
\vspace{-5mm}
\begin{proof}
Consider the following algorithm: The algorithm constructs another 
instance $\cal I'$ of the problem by multiplying the facility opening 
costs by $\alpha$, runs $\cal A$ on this modified instance $\cal I'$, 
and outputs its answer. It is easy to see that this algorithm is an 
LMP $\alpha$-approximation.
\end{proof}

\noindent Now we only need to prove the following. The proof of this theorem 
follows the general scheme that is explained in Section \ref{disc}.

\begin{theorem}
\label{F2}
For every instance $\cal I$ and every solution $SOL$ for $\cal I$,
Algorithm 2 finds a solution of cost at most $F_{SOL}+2 C_{SOL}$,
where $F_{SOL}$ and $C_{SOL}$ are facility and connection costs of $SOL$.
\end{theorem}
\vspace{-4mm}
\begin{proof}
By Theorem \ref{tradeoffthm} we only need to prove that the solution of the
factor-revealing LP \ref{tradeofflp} with $\gamma_f=1$ is at most 2. We first
write the maximization program \ref{tradeofflp} as the following
equivalent linear program.
\mathindent 0.5em
\begin{lp}
\label{simplelp} \maximize & {\sum_{i=1}^k\alpha_i - f}\\[\lpskip]
\st   & \sum_{i=1}^kd_i=1\nonumber\\
      & \forall\,1\le i<k:~         \alpha_i-\alpha_{i+1}\le0\nonumber \\
      & \forall\,1\le j<i< k:~      r_{j,i+1}-r_{j,i}\le0 \nonumber \\
      & \forall\,1\le j<i\le k:~    \alpha_i-r_{j,i}-d_i-d_j\le0 \nonumber \\
      & \forall\,1\le j<i\le k:~    r_{j,i}-d_i-g_{i,j}\le0\nonumber \\
      & \forall\,1\le i\le j\le k:~ \alpha_i-d_j-h_{i,j}\le0\nonumber \\[-1mm]
      & \forall\,1\le i\le k:~      \sum_{j=1}^{i-1}g_{i,j}+\sum_{j=i}^k h_{i,j}-f\le0 \nonumber\\
      & \forall\,i, j:~ \alpha_j, d_j, f,  r_{j,i}, g_{i,j}, h_{i,j}\ge0  \nonumber
\end{lp}
\mathindent 3em
We need to prove an upper bound of 2 on the solution of the above LP.
Since this program is a maximization program, it is enough to prove the upper
bound for any relaxation of the above program. Numerical results (for a fixed
value of $k$, say $k=100$) suggest that removing the second, third, and seventh 
inequalities of the above program does not change its solution. Therefore, we can
relax the above program by removing these inequalities. Now, it is a simple
exercise to write down the dual of the relaxed linear program and compute
its optimal solution. This solution corresponds to multiplying the 
third, fourth, fifth, and sixth inequalities of the linear program \ref{simplelp}
by $1/k$, and the first one by $(2-1/k)$, and adding up these inequalities.
This gives an upper bound of $2-1/k$ on the value of the objective function.
Thus, for $\gamma_f=1$, we have $\gamma_c\le2$. In fact, $\gamma_c$ is precisely
equal to $2$, as shown by the following solution for the program \ref{tradeofflp}.
\begin{eqnarray*}
\alpha_i&=&\left\{\begin{array}{ll} 2-\frac1k & i = 1 \\ 2 & 2\le
i\le k\end{array}\right.\\ d_i&=&\left\{\begin{array}{ll} 1 & i =
1 \\ 0 & 2\le i\le k\end{array}\right.\\
r_{j,i}&=&\left\{\begin{array}{ll} 1 & j = 1 \\ 2 & 2\le j\le
k\end{array}\right.\\ f&=&2(k-1)
\end{eqnarray*}
This example shows that the above analysis of the factor-revealing LP is tight.
\end{proof}

Lemma \ref{scalinglemma} and Theorem \ref{F2} provide an LMP $2$-approximation
algorithm for the metric facility location problem. This result improves
all the results in Jain and Vazirani~\cite{JV}, and gives straightforward
algorithms for some other problems considered by Charikar et al~\cite{CKMN}.

Notice that Theorem \ref{lmphard} shows that finding an LMP $(1+\frac2e-\ep)$-approximation
for the metric facility location problem is hard. Also, the integrality gap
examples found by Guha~\cite{guhathesis} show that Lemma \ref{kmedian}
is tight. This shows that one cannot use Lemma \ref{kmedian} as
a black box to obtain a smaller factor than $2+\frac4e$ for $k$-median problem.
Note that $3+\epsilon$ approximation is already known~\cite{kmed3}
for the problem. Hence if one wants to beat this factor using the
Lagrangian relaxation technique then it will be necessary to look
into the underlying LMP algorithm as already been done by 
Charikar and Guha~\cite{CG}.

\subsection{Facility location game}
An important consideration, in cooperative game theory, while distributing the
cost of a shared utility, is that the cost shares should satisfy the
{\em coalition participation constraint}, i.e., the total cost share of 
any subset of the users shall not be larger than their stand-alone cost of
receiving the service, so as to prevent this subset from seceding. In general,
this turns out to be a stringent condition to satisfy. 
For the facility location problem, Goemans and Skutella~\cite{GS}
showed that such a cost allocation is only possible for a very special case.
Furthermore, intractability sets in as well, for instance, in the case of
the facility location problem, computing the optimal cost of serving a
set of users is \NP-hard. 

In~\cite{JVgame} Jain and Vazirani relax this notion: for a constant $k$,
ensure that the cost share of any subset is no more than $k$ times its
stand-alone cost. They also observe that LP-based approximation algorithms 
directly yield a cost sharing method compatible with this relaxed notion. 
However, this involves solving
an LP, as in the case of LP-rounding. We observe that our facility location
algorithms automatically yield such a cost sharing method, with $k = 1.861$
and $k = 1.61$ respectively, 
by defining the cost share of city $j$ to be $\alpha_j$.

\subsection{Arbitrary demands}
In this version, for each city $j$, a non-negative integer demand $d_j$, is 
specified. An open facility $i$ can serve this demand at the cost of $c_{ij}d_j$. 
The best way to look at this modification is to reduce it to unit demand case by 
making $d_j$ copies of city $j$. This reduction suggests that we need to change 
our algorithms , so that each city $j$ raises its contribution $\alpha_j$ at rate $d_j$. 
Note that the modified algorithms still have the same running time in more general cases, 
where $d_j$ is fractional or exponentially large, and achieve the same approximation 
ratio.

\subsection{Fault tolerant facility location with uniform connectivity requirements}
We are given a connectivity requirement $r_j$ for each city $j$, which 
specifies the number of open facilities that city $j$ should be connected to. 
We can see that this problem is closely related to the set 
multi-cover problem, in the case that every set can be picked at most 
once~\cite{multicover}. 
The greedy algorithm for set-cover can be adapted for this variant of
the multi-cover problem achieving the same approximation factor.  We can 
use the same approach to deal with the fault tolerant facility location:
The mechanism of raising dual variables and opening facilities is the 
same as in our initial algorithms. The only difference is that city $j$ stops 
raising its dual variable and withdraws its contribution from other facilities,
when it is connected to $r_j$ open facilities. We can show that when all $r_j$'s 
are equal, our algorithms can still achieve the approximation factor of
1.861
and 1.61.  

\subsection{Facility location with penalties}
In this version we are not required to connect every city to an open 
facility; however, for each city $j$, there is a specified penalty, $p_j$, 
which we have to pay, if it is not connected to any open facility. We can modify
our algorithms for this problem as follows: If $\alpha_j$ 
reaches $p_j$ before $j$ is connected to any open facility, the city $j$ stops raising 
its dual variable and keeps its contribution equal to its penalty until it is either 
connected to an open facility or all remaining cities stop raising their dual variables.
At this point, the algorithm terminates and unconnected cities remain unconnected. 
Using the linear programming formulation introduced in Charikar~et al. (\cite{CKMN}
inequalities~(4.6)-(4.10)), we can show that the 
approximation ratio and running time of our modified algorithms have not
changed. 

\subsection{Robust facility location}
In this variant, we are given a number $l$ and we are only required to connect 
$n_c - l$ cities to open facilities. This problem can be reduced to the 
previous one via Lagrangian relaxation. Very recently, 
Charikar~et al.~\cite{CKMN} proposed a primal-dual 
algorithm, based on JV algorithm, which achieves an approximation ratio of 3. 
As they showed, the linear programming formulation of this variant has an 
unbounded integrality gap. In order to fix this problem,
they use the technique of parametric pruning, in which they guess the 
most expensive facility in the optimal solution. After that, they run JV algorithm
on the pruned instance, where the only allowable facilities are 
those that are not more expensive than the guessed facility. Here 
we can use the same idea, using Algorithm 1 rather than the JV algorithm.
Using a proof similar to the proof of the Theorem~3.2 in~\cite{CKMN}, 
we can prove that this algorithm solves the robust
facility location problem with an approximation factor of 2.

\subsection{Dealing with capacities}
In real applications, it is not usually the case that the cost
of opening a facility is independent of the number of cities
it will serve. But we can assume that we have {\em economy of scales},
i.e., the cost of serving each city decreases when the number
of cities increases (since publication of the first draft of this 
paper, this problem has also been studied in~\cite{HMM}). 
In order to capture this property, we define
the following variant of the capacitated metric facility location problem.
For each facility $i$, there is an initial opening cost $f_i$.
After facility $i$ is opened, it will cost $s_i$ to serve each city.
This variant can be solved using metric uncapacitated facility
location problem: We just have to change the metric such that
for each city $j$ and facility $i$, $c'_{ij} = c_{ij} + s_i$.
Clearly, $c'$ is also a metric and the solution of the metric
uncapacitated version to this problem can be interpreted as a
solution to the original problem with the same cost. 

We can reduce the variant of the capacitated facility location
problem in which each facility can be opened many times~\cite{JV}
to this problem by defining $s_i = f_i / u_i$. If
in the solution to this problem $k$ cities are connected to
facility $i$, we open this facility $\lceil k / u_i \rceil$ times.
The cost of the solution will be at most two times the original
cost so any $\alpha$-approximation for the uncapacitated
facility location problem can be turned into a $2\alpha$-approximation
for this variant of the capacitated version. We can also use the same
technique as in~\cite{JV} to give a factor $3$-approximation algorithm 
for this problem based on the LMP $2$-approximation algorithm for
uncapacitated facility location problem. 

\section{Discussion}
\label{disc}

The method of dual fitting can be seen as an implementation of the primal-dual
schema in which, instead of relaxing complementary
slackness conditions (which is the most common way of implementing the schema),
we relax feasibility of the dual. However, we prefer to reserve the term primal-dual
for algorithms that produce feasible primal and dual solutions.

Let us show how the combination of dual fitting with factor-revealing LP
applies to the set cover problem. The duality-based restatement of the greedy
algorithm (see~\cite{Va.book}) is:
All elements in the universal set $U$
increase their dual variables uniformly. Each element contributes its dual towards
paying for the cost of each of the sets it is contained in.
When the total contribution offered to a set equals its cost, 
the set is picked. At this point, the newly covered elements freeze their
dual variables and withdraw 
their contributions from all other sets. As stated in the introduction, the latter
(important) step ensures that the primal is fully paid for by the dual.
However, we might not get
a feasible dual solution. To make the dual solution feasible we look for 
the smallest positive number $Z$, so that when the dual solution is shrunk 
by a factor of $Z$, it becomes feasible. An upper bound on the approximation factor
of the algorithm is obtained by maximizing $Z$ over all possible
instances. 

Clearly $Z$ is also the maximum factor by which any set is
over-tight. Consider any set $S$. We want to see what is the worst
factor, over all sets and over all possible instances of the
problem, by which a set $S$ is over-tight. Let the elements in $S$
be $1,2, \cdots, k$. Let $x_i$ be the dual variable corresponding
to the element $i$ at the end of the algorithm. Without loss of 
generality we may assume that $x_1 \leq x_2\le \cdots\le x_k$. 
It is easy to see that at time $t=x_i^-$, total duals offered 
to $S$ is at least $(k-i+1)x_i$. Therefore, this value cannot
be greater than the cost of the set $S$ (denoted by $c_S$).
So, the optimum solution of the following mathematical program
gives an upper bound on the value of $Z$. 
(Note that $c_S$ is a variable not a constant).

\begin{lp}
\label{lp1} \maximize & \frac{\sum_{i=1}^k x_i}{c_S} \\[\lpskip]
\st & \forall 1\leq i < k :~ x_i \le x_{i+1} \nonumber \\
      & \forall 1\leq i \leq k:~ (k-i+1) x_i \leq c_S  \nonumber \\
      & \forall 1\leq i \leq k:~ x_i \geq 0  \nonumber \\
      & c_S \ge 1 \nonumber
\end{lp}

The above optimization program can be turned into a linear program by
adding the constraint $c_S=1$ and changing the objective function to
$\sum_{i=1}^k x_i$. We call this linear program the {\em factor-revealing LP}. 
Notice that the factor-revealing LP has nothing to do with
the LP formulation of the set cover problem; it is only used in 
order to analyze this particular algorithm. This is the important 
distinction between the factor-revealing LP technique, and other LP-based 
techniques in approximation algorithms.

One advantage of reducing the analysis of the approximation guarantee of
an algorithm to obtaining an upper bound on the optimal solution to
a factor-revealing LP is that one can introduce emperical experimentation
into the latter task. This can also help decide which aspects of
the execution of the algorithm to introduce into the factor-revealing LP
to obtain the best possible bound on the performance of the algorithm,
e.g., we needed to introduce the variables 
$r_{j,i}$ in Section \ref{factorlpsec} in order to get a good bound
on the approximation ratio of Algorithm 2. 

In general, this technique is not guaranteed to yield a tight analysis of
the algorithm, since the algorithm may be performing 
well not because of local reasons but for some global reasons that 
are difficult to capture in a factor-revealing LP. In the case of set cover,
this method not only produces a tight analysis, but the factor-revealing LP
also helps produce a tight example for the algorithm. From any feasible
solution $x$ of factor-revealing LP \ref{lp1}, one can construct 
the following instance: There are $k$ elements $1,\ldots,k$,
a set $S=\{1,\ldots,k\}$ of cost $1+\ep$ which is the optimal solution, and
sets $S_i=\{i\}$ of cost $x_i$ for $i=1,\ldots,k$. It is easy to verify
that the greedy algorithm gives a solution that is $\sum x_i$ times 
worse than the optimal on this instance. Picking $x$ to be the optimal
solution, we get a tight example, and also show that the approximation ratio of the 
greedy algorithm is precisely equal $H_n$, the optimal solution 
of the factor-revealing LP.

Finally, in terms of practical impact,
what is the significance of improving the approximation guarantee
for facility location from 3 to 1.81 or 1.61 when practitioners are
seeking algorithms that come within 2\% to 5\% of the optimal?
The superior experimental results of our algorithms, as compared with the
JV algorithm, seem to provide the answer and to support the
argument made in~\cite{Va.book} (Preface, page IX) that the approximation factor
should be viewed as a ``measure that forces us to explore
deeper into the combinatorial structure of the problem and discover more
powerful tools for exploiting this structure'' and the observation that
``sophisticated algorithms do have the error bounds of the desired magnitude,
2\% to 5\%, on typical instances, even though their worst case error
bounds are much higher''.

\noindent {\bf Acknowledgments.}
We would like to thank Michel Goemans, Mohammad Ghodsi, Nicole Immorlica, 
Nisheet K. Vishnoi, Milena Mihail, and Christos Gkantsidis for their helpful comments and discussions.

\bibliography{flp}
\bibliographystyle{plain}
\end{document}